\documentclass[11pt]{article}
\usepackage{amsmath, amssymb, geometry, setspace}
\usepackage{float}
\usepackage{booktabs}
\usepackage{parskip}
\usepackage{graphicx}
\usepackage{xcolor}
\usepackage[colorlinks=true, linkcolor=blue, citecolor=blue, urlcolor=blue]{hyperref}
\usepackage{natbib}
\title{Powerful Switchback Experiments -- Or Not?}
\author{Sergei Pankratev \\ DoorDash, Inc.}
\date{}

\begin{document}
\maketitle

\begin{abstract}
Switchback experiments---in which treatment is assigned at the level of a cluster crossed with a time period---are widely used in marketplace and platform settings, yet no closed-form power formula exists for them.
We fill this gap by deriving a closed-form, multi-level asymptotic variance approximation for the individual-level OLS estimator, facilitating power budgeting. 
Using this formula, we reveal a structural floor on statistical power: while idiosyncratic noise vanishes with observation density, macro-level shocks are multiplicatively penalized by cluster size imbalance. 
We confirm through analytical derivations and Monte Carlo simulations that the formula is exact across typical parameters and serves as a mathematically conservative upper bound in extreme boundary regimes.
We study three methodological applications. 
First, we prove that advanced assignment designs like stratification only partially eliminate the penalty of cluster size imbalance on power. 
Second, we demonstrate that variance reduction techniques targeting macro-level shocks yield disproportionately greater efficiency gains than those targeting residual noise. 
Third, we formalize the finite-sample power trade-offs between individual-level and cell-level estimators.
\end{abstract}

\section{Introduction}

Switchback experiments are a natural design for platforms and marketplaces in which the unit of business interest is a cross-sectional entity---such as a geographic region, a service provider, or a network of users---observed repeatedly over time.\footnote{
    We adopt the following terminology throughout. 
    A \textit{cluster} is a cross-sectional entity (e.g., a delivery zone or a user network), and a \textit{cell} is a (cluster $\times$ time period) combination that serves as the unit of randomization. 
    The term \textit{unit} refers to an individual observation within a cell.
}
This design is widely used at firms such as DoorDash, Lyft, Airbnb, and other platform companies \citep{bajari2021multiple, bojinov2023design, hu2022average, johari2022experimental}, precisely because randomizing treatments over time within the same market mitigates the network effects that bias standard A/B tests \citep{kohavi2020trustworthy}.
Despite their prevalence, estimating power in switchback experiments correctly remains a challenge. 
Furthermore, the structure of switchback experiments introduces nuances to the application of standard experimental design and analysis strategies---such as stratification, variance reduction, and estimator choice---that remain largely unexplored.

To address this gap, we derive a closed-form analytical variance approximation that accommodates the full structure of a switchback design.
Specifically, we consider a setting with $J$ clusters and $H$ time periods, where each cluster contains $n_{j,h}$ observations in period $h$, and error components are decomposed into cluster effects, temporal effects, cluster-temporal interactions, and idiosyncratic residuals.
Applying a Delta method linearization to the individual-level OLS estimator, we obtain a closed-form approximation for its variance.
We analytically derive the approximation's boundary conditions to establish its finite-sample reliability. 
Finally, we demonstrate how the formula guides practical experimental design by formalizing the power trade-offs between estimators and optimizing the deployment of variance reduction techniques.

We validate the formula through extensive Monte Carlo simulations, proving that it tracks empirical variance across typical industry parameters. 
Furthermore, we explicitly map the boundary conditions where the approximation breaks down (e.g., highly skewed cluster sizes $cv \ge 2.0$ or $J \times H \le 10$). 
In these extreme edge cases, our analytical formula smoothly transitions into a mathematically conservative upper bound.

% Results & Why this matters
Our core contribution is a seemingly small methodological step---deriving a single closed-form asymptotic variance formula---but it resolves a significant practical bottleneck in industry. 
This is not merely a convenience gap. 
By providing an explicit method for budgeting power, the formula allows practitioners to evaluate design trade-offs without relying on computationally expensive Monte Carlo simulations. 
Beyond its primary utility for power budgeting, the framework yields four key insights for experimental design.

First, it reveals a structural floor on statistical power caused by the \textit{imbalance penalty}. 
While the idiosyncratic residual variance component diminishes predictably with observation density, the variance contribution of macro-level shocks---systematic perturbations affecting a cluster, a time period, or their intersection---is multiplicatively penalized by the factor $(1 + cv^2)$, where $cv$ is the coefficient of variation of cluster sizes. 
This multi-level penalty functions as a modern generalization of the classic Moulton factor \citep{moulton1986random}. 
This implies that increasing the density of observations within existing cells yields rapidly diminishing returns for power.
% Conversely, we prove that temporal autocorrelation drops out entirely when randomizing at the cell level, averting what would otherwise be a severe variance inflation.

Second, the formula quantitatively evaluates the theoretical limits of advanced treatment assignment designs. 
We prove that while methods like within-cluster stratification and pairing mitigate specific spatial and temporal variance penalties, they leave interaction variance fully exposed to cluster imbalance, maintaining a strict upper bound on their efficiency.

Third, the formula guides the strategic deployment of variance reduction techniques. 
We show that predicting idiosyncratic shocks may have a relatively minor impact on variance compared to predicting macro-level shocks, as the latter are actively amplified by cluster size imbalance.

Fourth, the formula formalizes the exact mathematical trade-off between individual-level and cell-level estimators. 
We demonstrate that while individual-level estimators suffer from the $(1+cv^2)$ cluster imbalance penalty, cell-level estimators suffer from low cell density by forcing equal weight onto highly volatile low-density cells. 
This provides a rigorous framework for analysts to choose their aggregation level based on their specific platform's cell density and cluster size imbalance.

The paper is organized as follows.
Section~\ref{sec:literature} provides a review of related literature.
Section~\ref{sec:model} formalizes the theoretical framework, including the model, the variance formula, and its boundary conditions.
Section~\ref{sec:practical} discusses practical considerations, including estimator choice and the value of variance reduction.
Section~\ref{sec:simulations} presents the simulation validation.

% The appendices contain the full derivation of the variance formula (\hyperref[app:derivation]{Appendix~\ref*{app:derivation}}), the formal proof of the finite-sample covariance breakdown (\hyperref[app:boundary]{Appendix~\ref*{app:boundary}}), the derivation of the cell-level variance (\hyperref[app:bucket]{Appendix~\ref*{app:bucket}}), and a summary of approximation errors across parameter regimes (\hyperref[app:sim_errors]{Appendix~\ref*{app:sim_errors}}).

\section{Related Literature}
\label{sec:literature}

% \paragraph{Switchback experiment design and inference.}
Switchback experiments have received growing methodological attention driven by their deployment at large platform firms.
\citet{bojinov2023design} establish a formal potential outcomes framework for switchback designs and derive conditions under which the difference-in-means estimator is unbiased for the average treatment effect under no-interference and finite-population asymptotics; their analysis focuses on identification and the structure of treatment carry-over.
Recent literature explores complex design modifications to improve the efficiency of switchbacks, such as dynamic spatial clustering \citep{ni2023design} or rerandomization \citep{ni2025reliable}. 
However, the standard switchback design remains widely used in industry due to its operational feasibility and robustness to various sources of interference that can occur in multisided marketplaces. 
Our work provides the missing analytical foundation for this standard design, empowering practitioners to properly budget power and quantify the impact of structural constraints like cluster imbalance.

Turning to variance estimation, \citet{liu2026randomization} develop randomization-based inference for switchbacks and provide a variance approximation. 
However, their setting assumes a single time series (one cross-sectional unit observed over multiple periods), which cannot capture the spatial heterogeneity and cluster-size imbalance central to marketplace experiments.

The most natural starting points for our result come instead from the cross-sectional literature on cluster-randomized trials (CRTs). 
The classical design effect derived by \citet{killip2004intracluster} shows how intra-cluster correlation inflates variance, but assumes that treatment is assigned permanently, temporal dynamics do not exist, and all clusters are exactly equal in size. 
\citet{eldridge2006sample} relax the equal-size assumption, demonstrating that cluster size imbalance amplifies the variance penalty by a multiplicative factor of $(1+cv^2)$, where $cv$ is the coefficient of variation of cluster sizes. 
Our formula extends this critical insight to the richer multi-level error structure of a switchback design, where cluster, temporal, interaction, and residual variance components are affected by imbalance in qualitatively different ways.

Interestingly, this severe multiplicative penalty on macro-level shocks distinguishes switchbacks from other longitudinal cluster-randomized designs.
For example, \citet{kristunas2017imbalance} demonstrate that in stepped-wedge trials---where treatment assignment monotonically ratchets in one direction rather than flipping back and forth---cluster size imbalance does not lead to a notable loss of power.
In a switchback, the continual reshuffling of treatment creates a dynamic where large, volatile clusters constantly destabilize the estimator's denominator, making the design uniquely vulnerable to imbalance.

\section{Theoretical Framework}
\label{sec:model}

\subsection{Setup and Estimator}
We consider an experiment with $J$ clusters (indexed $j = 1, \ldots, J$) and $H$ time periods (indexed $h = 1, \ldots, H$).
In each cell $(j, h)$, there are $n_{j,h} \geq 1$ individual observations.
Let $\bar{n} = \frac{1}{JH}\sum_{j,h} n_{j,h}$ denote the mean cell size and $cv = \mathrm{std}(n_{j,h})/\bar{n}$ the coefficient of variation (the ratio of the standard deviation to the mean) of cell sizes across all $JH$ cells.

The potential outcome for individual $i$ in cell $(j, h)$ under assignment $W$ is
\begin{equation}
\label{eq:dgp}
Y_{i,j,h}(W) = \mu + \tau W_{j,h} + \alpha_j + \gamma_h + \delta_{j,h} + \epsilon_{i,j,h},
\end{equation}
where $W_{j,h} $ is treatment randomly assigned at the cell level, $\tau$ is the treatment effect, and $(\alpha_j, \gamma_h, \delta_{j,h}, \epsilon_{i,j,h})$ are the four error components that are mutually independent and mean zero.
Specifically, $\alpha_j \sim \text{iid}(0, \sigma_{cl}^2)$ captures permanent between-cluster heterogeneity; $\gamma_h \sim \text{iid}(0, \sigma_{time}^2)$ captures time effects common to all clusters in period $h$; $\delta_{j,h} \sim \text{iid}(0, \sigma_{int}^2)$ captures cluster-specific deviations in period $h$; and $\epsilon_{i,j,h} \sim \text{iid}(0, \sigma_{res}^2)$ is within-cell idiosyncratic noise.

We define the total variance $\sigma_{total}^2 = \sigma_{cl}^2 + \sigma_{time}^2 + \sigma_{int}^2 + \sigma_{res}^2$ and variance shares $S_k = \sigma_k^2 / \sigma_{total}^2$ for $k \in \{cl, time, int, res\}$, so that $S_{cl} + S_{time} + S_{int} + S_{res} = 1$. For convenience, we define the macro-level variance share as $S_{macro} = S_{cl} + S_{time} + S_{int}$.

The individual-level Ordinary Least Squares (OLS) estimator for $\tau$ is obtained by regressing the outcome $Y_{i,j,h}$ on the treatment assignment $W_{j,h}$.
Throughout the paper, we refer to this estimator interchangeably as the \textit{individual-level} OLS estimator or the individual-level difference-in-means. 
This is in contrast to the \textit{cell-level} difference-in-means, which computes an unweighted t-test (or equivalently, an OLS regression) on the aggregated cell-level outcomes.

% Because the number of treated observations is random, the resulting OLS point estimate $\hat{\tau}$ behaves mathematically as a ratio estimator.\footnote{
%     Specifically, because treatment $W_{j,h}$ is randomized at the cell level, the total number of treated individuals, $N_T = \sum_{j,h} W_{j,h} n_{j,h}$, is a random variable determined by the specific realization of the assignment vector. 
%     Consequently, the individual-level difference-in-means is a ratio of two random variables (the random sum of treated outcomes divided by the random sum of treated individuals), rather than a simple average with a fixed denominator.
% }
% Its exact finite-sample distribution is analytically intractable, so we proceed via a Delta method approximation to derive the true asymptotic variance of this estimate.
 % COMMENT: for thiiiiis i should carefully read and check everything

\subsection{The Variance Formula}
\label{sec:variance}

The primary determinant of statistical power is the variance of the treatment effect estimator. 
To derive this variance, we first express the individual-level OLS estimate $\hat{\tau}$ as the difference of two ratios, since both the outcomes and the exact number of observations in each treatment arm are random variables.\footnote{
    Specifically, because treatment $W_{j,h}$ is randomized at the cell level, the total number of treated individuals, $N_T = \sum_{j,h} W_{j,h} n_{j,h}$, is a random variable determined by the specific realization of the assignment vector. 
    Consequently, the individual-level difference-in-means is a ratio of two random variables (the random sum of treated outcomes divided by the random sum of treated individuals), rather than a simple average with a fixed denominator.
}
We then linearize these ratios using a Delta method approximation to isolate the estimation error. 
Finally, we evaluate the variance of this linearized error across the multi-level outcome components defined in \eqref{eq:dgp}. 
This yields the following closed-form approximation for the asymptotic variance:

\begin{equation}
\label{eq:variance}
\text{Var}(\hat{\tau}) \approx \frac{4 \cdot \sigma_{total}^2}{J \cdot H} \left[ \frac{S_{res}}{\bar{n}} + S_{macro} \left( \frac{1}{\bar{n}} + 1 + cv^2 \right) \right]
\end{equation}

where $J$ is the number of clusters, $H$ is the duration, $\bar{n}$ is the mean cluster-hour density, $cv = \mathrm{std}(n_{j,h})/\bar{n}$ is the coefficient of variation of the cell sizes, and $S_k$ represents the variance shares of the respective components.\footnote{
    The formula derived here represents the expected value of the standard cluster-robust sandwich variance estimator (CRVE) applied to this data generating process.
}
We provide the full mathematical derivation in \hyperref[app:derivation]{Appendix~\ref*{app:derivation}}.

By rewriting the terms algebraically, we can show that the variance of a switchback experiment cleanly decomposes into the variance of a naive A/B test plus a specific penalty term. 
In a standard, completely randomized A/B test with $N = J \times H \times \bar{n}$ total observations, the variance is $\frac{4 \cdot \sigma_{total}^2}{N}$.
By expanding our formula, we find that the switchback variance decomposes into:

\begin{equation}
\label{eq:decomp}
\text{Var}(\hat{\tau}) \approx \underbrace{\frac{4 \cdot \sigma_{total}^2}{N}}_{\text{Naive A/B Variance}} + \underbrace{\frac{4 \cdot \sigma_{macro}^2}{J \cdot H} (1 + cv^2)}_{\text{Switchback Penalty}}
\end{equation}

This decomposition mathematically proves that a switchback experiment is strictly less powerful than a naive A/B test on the exact same data (\hyperref[app:decomp]{Appendix~\ref*{app:decomp}} provides the derivation).\footnote{
    While a naive individual-level A/B test is causally invalid in these settings due to network interference, it serves as the exact mathematical baseline from which the spatial-temporal design penalty diverges.
} 
This power loss is driven entirely by how the two different types of noise interact with the randomization design.

The first component of the right-hand side of \eqref{eq:variance}, idiosyncratic residual noise ($S_{res}$), behaves exactly as standard sampling theory predicts. 
It is scaled by $1/\bar{n}$, meaning that as the number of observations per cell grows, the idiosyncratic noise attenuates and its contribution to the overall variance vanishes.

In contrast, macro-level shocks ($S_{macro}$)---those occurring at the cluster level, hour level, or cluster-hour level---do not enjoy this complete averaging.
Because these shocks are shared by all $n_{j,h}$ observations in a given cell, they are scaled up by the cell size before entering the estimator numerator.
Consequently, they are multiplied by the factor $(1/\bar{n} + 1 + cv^2)$.
As the data density $\bar{n}$ increases, the $1/\bar{n}$ term approaches zero, but the $(1+cv^2)$ multiplier remains strictly bounded away from zero.

This $(1+cv^2)$ term is the fundamental penalty of cluster size imbalance.
Because treatment is assigned at the cell level, a random assignment that happens to place disproportionately large cells into the treatment arm will give substantial weight to the macro-level shocks of those specific cells. Since all individuals in a cell share the exact same cluster or temporal error, these errors do not average out. Instead, they are multiplied by the disproportionately large cell size, severely destabilizing the estimator.
This structural vulnerability scales with the squared coefficient of variation of the cell sizes.\footnote{
    We assume that the variance of interaction shocks ($\sigma^2_{int}$) is independent of cluster size. 
    If interaction shocks exhibit heteroskedasticity and scale positively with cluster size (e.g., larger clusters experience a wider distribution of interaction shocks), the $(1+cv^2)$ penalty on $S_{int}$ becomes even more severe than the baseline formula predicts.
}
Conversely, because treatment is re-randomized independently at every time period, any underlying temporal autocorrelation in the data is neutralized in expectation, averting what would otherwise be a severe variance inflation.

To quantify the magnitude of this power loss, consider a typical dense, imbalanced marketplace setting ($\bar{n} = 20$, $cv = 1.5$) where the raw variance is heavily dominated by idiosyncratic noise ($S_{res} = 80\%$, $S_{macro} = 20\%$). 
Factoring out the common multiplier $\frac{4 \cdot \sigma_{total}^2}{J \cdot H}$ from \eqref{eq:decomp}, the naive A/B testing variance is proportional to $1/\bar{n} = 0.05$. 
However, the switchback penalty is proportional to $S_{macro}(1 + cv^2) = 0.20 \times 3.25 = 0.65$. 
Thus, despite macro-level shocks accounting for only $20\%$ of the raw noise, the cluster imbalance inflates the total variance to $0.70$. 
A switchback experiment in this setting requires exactly 14 times more data ($0.70 / 0.05$) to achieve the same statistical power as a naive A/B test.

This reveals a fundamental constraint on experimental design in marketplaces: accumulating a substantial volume of observations within existing clusters (large $\bar{n}$) yields rapidly diminishing returns for statistical power. 
As long as cluster imbalance ($cv$) remains high and macro-shocks are prevalent, the variance of the estimator will not vanish.
To increase power, a practitioner must either increase the number of independent randomizations (by increasing the number of clusters $J$ or the experiment duration $H$), reduce the imbalance itself, or use variance reduction techniques to explicitly shrink the macro-level variance shares.

Because the individual-level OLS point estimate is asymptotically normal, this variance approximation can be plugged directly into standard minimum detectable effect calculations ($\mathrm{MDE} = (z_{1-\alpha/2} + z_\beta) \times \sqrt{\text{Var}(\hat{\tau})}$) and sample-size budgeting. 
We provide the explicit, expanded MDE and required sample-size formulas for practitioners in \hyperref[app:mde]{Appendix~\ref*{app:mde}}.

\subsection{Special Cases}
Our formula cleanly collapses to established results in simpler settings.
For a standard cluster-randomized trial with a single time period ($H=1$), no time effects ($S_{time} = S_{int} = 0$), and balanced clusters ($cv=0$, $n_{j,h} = m$), the variance simplifies to being proportional to $\frac{1}{J} (S_{cl} + S_{res}/m)$, recovering the classical \citet{killip2004intracluster} design effect $\mathrm{DE} = 1 + \rho(m-1)$.
If we allow cluster sizes to vary ($cv > 0$) under these same conditions, the formula recovers the exact $(1 + cv^2)$ correction derived by \citet{eldridge2006sample} for unequal cluster sizes.\footnote{
    Our formula does not collapse to the time-series approximation derived by \citet{liu2026randomization}. 
    Their setting assumes a single cross-sectional unit observed over time ($J=1$). 
    In our spatial-first framework, setting $J=1$ breaks the core assumption that the cluster shock ($\alpha_j$) is a random variable distributed across multiple independent units; with only one cluster, the cluster effect becomes a constant baseline that cancels out in expectation. 
    Furthermore, their analysis focuses on temporal autocorrelation under alternating treatment sequences, whereas our cell-level Bernoulli assignment mathematically neutralizes temporal covariance entirely.
}

\section{Approximation Boundaries and Simulation Validation}
\label{sec:simulations}

In this section, we analyze the theoretical boundaries of the variance formula approximation and validate it using Monte Carlo simulations.

The analytical formula relies on a first-order Taylor expansion (the Delta method). While accurate for typical parameters, it systematically overpredicts variance in highly skewed or sparse finite-sample regimes. This occurs because the first-order approximation fixes the sample size at its expectation, ignoring the strong positive covariance between the experiment's squared error and realized sample size. Specifically, when a disproportionately large cluster enters the treatment group, both values increase substantially. Because the approximation ignores this correlation, it systematically overestimates the variance (we formally prove this breakdown in \hyperref[app:boundary]{Appendix~\ref*{app:boundary}}). Consequently, in these boundary regimes, our formula safely acts as a mathematically conservative upper bound.

To validate the analytical formula across the main parameter space and empirically map the exact boundaries of this predicted breakdown, we compare its predictions against empirical variances generated via Monte Carlo simulations. 
We refer to the skewed or sparse scenarios where the approximation breaks down as the \textit{boundary regime}, and the vast remaining parameter space where the approximation performs reliably as the \textit{interior regime}.

% \paragraph{Simulation Setup.}
We use a Data Generating Process (DGP) that exactly matches the model in \eqref{eq:dgp}.\footnote{
    The data generating process abstracts away deterministic trends along any of the experimental dimensions. 
    For instance, consider diurnal seasonality: because treatment assignment $W_{j,h}$ varies across time periods, any deterministic temporal trend is orthogonal to treatment in expectation and acts as an additional source of baseline variation, absorbing mathematically into the temporal variance component $\sigma_{time}^2$. 
    This same logic applies to deterministic cross-sectional or interaction trends, which similarly absorb into their respective variance components without altering the structural form of the variance approximation.
}
We establish a realistic baseline configuration and vary each parameter individually to isolate its specific effect on the variance, as detailed in Table~\ref{tab:sim_params}. 

\begin{table}[htbp]
\centering
\caption{Simulation baseline and parameter sweeps.}\label{tab:sim_params}
\renewcommand{\arraystretch}{1.20}
\resizebox{0.7\linewidth}{!}{%
\begin{tabular}{@{} l l l @{}}
    \toprule
    \textbf{Parameter} & \textbf{Baseline} & \textbf{Swept values} \\
    \midrule
    Number of clusters ($J$) & 100 & $\{10,\, 20,\, 50,\, 100,\, 250,\, 500\}$ \\
    Hours ($H$) & 168 & $\{24,\, 48,\, 168,\, 336,\, 720\}$ \\
    Cluster-size CV ($cv$) & 1.0 & $\{0.0,\, 0.5,\, 1.0,\, 2.0,\, 4.0\}$ \\
    Mean cell density ($\bar{n}$) & 20 & $\{2,\, 5,\, 10,\, 20,\, 50,\, 100\}$ \\
    Residual variance share ($S_{res}$) & 0.70 & $\{0.25,\, 0.50,\, 0.70,\, 0.85\}$ \\
    Autocorrelation ($\rho$) & 0.3 & $\{0.0,\, 0.3,\, 0.6,\, 0.9\}$ \\
    Total variance scale ($\sigma_{total}$) & $1{,}000$ & $\{500,\, 1{,}000,\, 2{,}000\}$ \\
    \bottomrule
\end{tabular}}
\end{table}

For each parameter regime, we simulate $1{,}000$ independent replications for lower-variance regimes, and $5{,}000$ replications for high-variance regimes.\footnote{
    Specifically, we increase to $5{,}000$ replications for regimes exhibiting high estimator kurtosis: severe cluster imbalance ($cv \ge 2.0$), exceptionally few clusters ($J \le 20$), or a dominant macro-level variance share ($S_{macro} \ge 0.50$, i.e., $S_{res} \le 0.50$). 
    In these cases, heavy-tailed distributions inflate the Monte Carlo error of the empirical variance and necessitate a larger sample to achieve stability.
}
In each replication, we draw a new dataset, apply individual-level random assignment ($W_{j,h} \sim \mathrm{Bernoulli}(0.5)$), and compute the OLS estimate $\hat{\tau}$.
The ``empirical variance'' is the variance of these estimates across the replications.

\begin{figure}[htbp]
    \centering
    \includegraphics[width=0.5\textwidth]{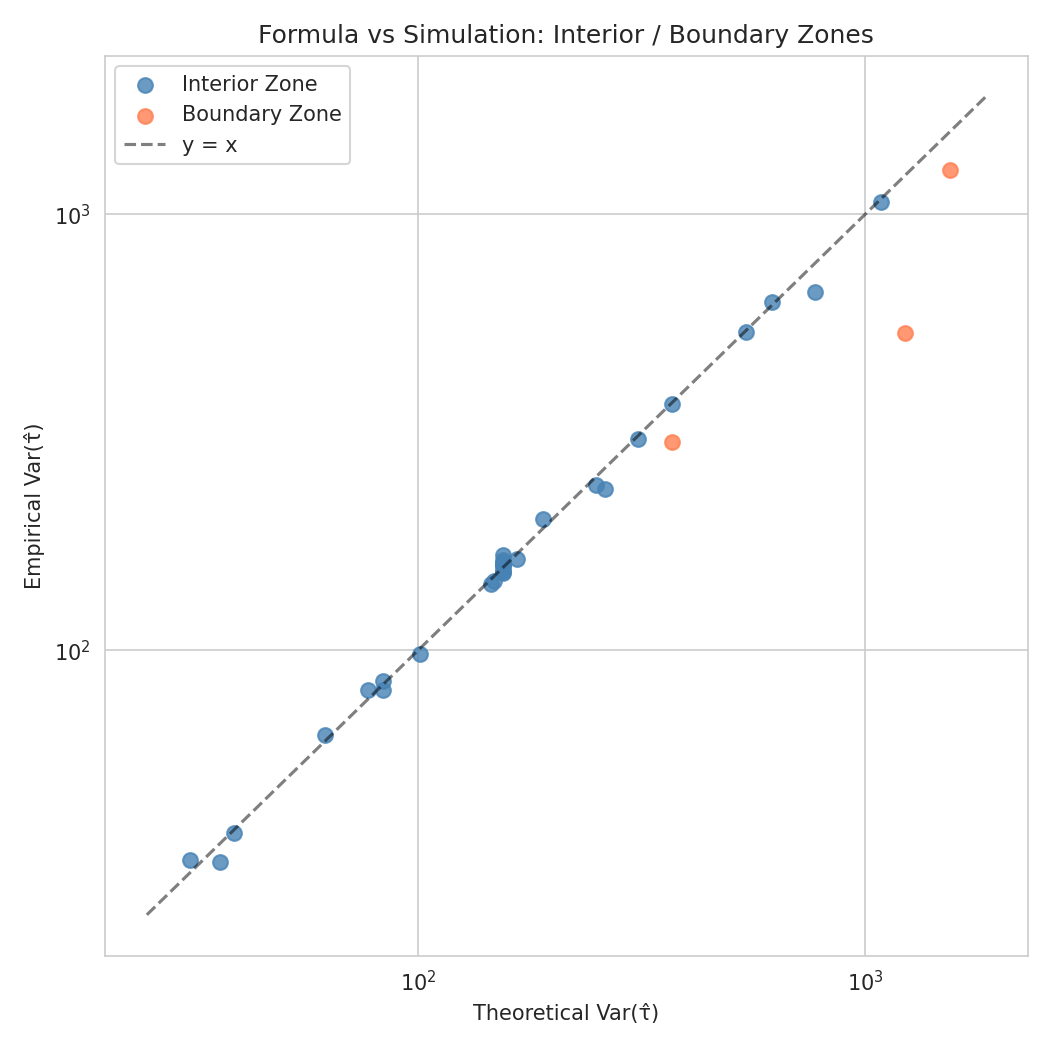}
    \caption{Theoretical versus Empirical Variance.}
    \label{fig:scatter}
\end{figure}

As illustrated by Figure~\ref{fig:scatter}, the formula closely tracks the empirical variance across four orders of magnitude of scale. In this figure, blue points represent the Interior regimes, while orange points represent the Boundary regimes.
For Interior regimes, the mean approximation error remains tightly bounded between 1.78\% and 6.57\%. 
We provide a comprehensive summary table of these errors, alongside further empirical validation of the formula's structural mechanics, in \hyperref[app:sim_errors]{Appendix~\ref*{app:sim_errors}}.
Across the vast majority of our parameter grid, the closed-form approximation is essentially exact.

Our simulations isolate the precise numerical boundaries where this breakdown occurs. 
Through these Monte Carlo simulations, we empirically bound this divergence to regimes where cluster sizes are highly skewed ($cv \geq 2.0$) or the total number of independent randomizations is exceptionally small ($J \times H \leq 10$).
In these regimes (highlighted in orange in our figures), the formula systematically overpredicts the empirical variance, precisely as the theoretical intuition predicts, with mean errors exceeding 15-25\%.
Additionally, a separate, minor divergence emerges at extremely low data densities ($\bar{n} \le 5$), although the resulting approximation error remains substantially lower than the severe overprediction caused by high cluster imbalance.

\section{Applications}
\label{sec:practical}

Our decomposition of the treatment effect variance for switchback experiments sheds light on several methodological aspects, including experimental design, variance reduction, and estimator choice. 
In this section, we apply our analytical framework to explore these nuances.

\subsection{Treatment Assignment Designs}
\label{sec:stratification}

A common heuristic to improve the efficiency of switchback experiments is to constrain the randomization using advanced operational designs. 
Standard stratification approaches include within-cluster stratification, pairing of similar clusters, and mirroring (which combines both).\footnote{These terms refer respectively to temporal blocking, cross-sectional matching, and matched crossover designs; for formal definitions and properties, see \citet{bojinov2023design} and \citet{ni2023design}.} 
By applying our variance decomposition, we can precisely quantify how these designs mitigate the switchback penalty.

These methods partially mitigate the cluster imbalance penalty by balancing assignments along different dimensions. 
Within-cluster stratification balances time for a single location: it ensures each cluster spends exactly half its time in the treatment condition, effectively reducing the penalty on the spatial variance ($S_{cl}$). 
Conversely, pairing balances space for a single time period: it matches clusters of similar sizes and assigns them opposite treatments, reducing the penalty on the temporal variance ($S_{time}$).\footnote{
    % \textcolor{blue}{
    In practice, experimenters often match clusters not only on size but also on historical baseline outcomes. 
    This induces a positive correlation between the spatial errors ($\alpha_j$) of the paired clusters, which actively reduces the spatial variance penalty ($S_{cl}$) beyond what is captured by our baseline i.i.d. assumption. 
    However, as we show below, this does not alter the fundamental limitation regarding interaction variance.
    % }
} 
Finally, mirroring applies both stratification and pairing simultaneously, protecting both the spatial and temporal variance components from the cluster imbalance multiplier.

However, none of these approaches addresses the interaction component ($S_{int}$) of the macro noise. 
Because interaction shocks ($\delta_{j,h}$) represent unpredictable, transient spatiotemporal deviations, our derivation proves that no \textit{fixed} pre-assignment design (such as classical pairing or mirroring) can balance them in expectation. 
The $(1+cv^2)$ multiplier permanently attaches to the interaction variance share under static assignment (we provide the full formal algebraic proofs in \hyperref[app:stratification]{Appendix~\ref*{app:stratification}}). 
Mitigating this specific variance component requires either post-experimental adjustment via machine learning models (e.g., CUPAC) that capture real-time spatial features, or the adoption of dynamic, adaptive experimental designs such as dynamic spatial clustering \citep{ni2023design} or sequentially-rerandomized switchbacks \citep{zeng2026sequentially}.

\subsection{Variance Reduction}
In practice, experimenters often employ variance reduction techniques, such as CUPED \citep{deng2013improving} or CUPAC \citep{poyarkov2016boosted}, to improve power.\footnote{
    While we focus on covariate adjustment methods like CUPED and machine-learning-based CUPAC, variance can also be computationally reduced via alternative methods, such as doubly robust estimators \citep{chernozhukov2018double}.
}
These methods effectively shrink specific components of the baseline outcome variance prior to the experiment. 
Our formula sheds light on important nuances that the structure of switchback experiments introduces to the application of variance reduction methods.

In a standard A/B test, reducing individual-level noise is highly advantageous. 
Consequently, complex machine-learning models (CUPAC) that accurately predict idiosyncratic user behavior often substantially outperform simple historical cell averages (CUPED). 
However, in a switchback experiment, the idiosyncratic residual noise is naturally suppressed by the $1/\bar{n}$ density factor, whereas macro-level noise is actively amplified by the $(1+cv^2)$ cluster imbalance penalty. 
Because of this structural inversion, the relative advantage of predicting individual-level noise diminishes severely. 
When implementing CUPAC for a switchback experiment, it is critical that the model is explicitly engineered to capture macro-level spatial and temporal features, rather than predicting individual-level residuals. 

To illustrate why targeting macro-level noise is so vital, suppose a platform's raw variance is heavily dominated by idiosyncratic noise ($S_{res} = 80\%$) rather than macro-level shocks ($S_{macro} = 20\%$). 
A naive view suggests that reducing this large residual share $S_{res}$ would be most effective.
However, the absolute variance reduction ($\Delta V$) depends on the structural multipliers. 
In a dense, imbalanced marketplace ($\bar{n} = 20$, $cv = 1.5$), halving either component yields substantially different results. 
Targeting macro shocks yields $\Delta V \propto 0.50 \times S_{macro} \times (1 + cv^2) = 0.50 \times 0.20 \times 3.25 = 0.325$, whereas targeting the residual component yields $\Delta V \propto 0.50 \times S_{res} \times (1/\bar{n}) = 0.50 \times 0.80 \times 0.05 = 0.02$.

Despite targeting a variance component four times smaller in the raw data, adjusting for macro-level noise is 16.25 times more effective. 
Because the residual share is structurally divided by $\bar{n}$, reducing it yields minor efficiency gains, whereas the $(1+cv^2)$ amplified macro share provides an extremely high-leverage target for variance reduction.
% \footnote{
%     We structure this comparison purely for depictive purposes to highlight the structural multipliers. 
%     In reality, modern ML-based CUPAC models may include both pre-experimental and real-time covariates, allowing them to capture substantial macro-level spatial and temporal variance alongside the residual noise.
% }

\subsection{Estimator Level Choice}
Our derivation relies on the individual difference-in-means estimator weighting cells proportionally to their size.
If an analyst instead aggregates the data into $(j,h)$ cells and computes an unweighted cell-level difference-in-means across these $J \times H$ cells, the variance properties change fundamentally.

Because the cell-level difference-in-means estimator is an unweighted average of independent assignments, its variance collapses precisely to the standard A/B testing variance formula: $\text{Var}(\hat{\tau}_{cell}) \approx \frac{4}{JH}\text{Var}(\bar{Y}_{j,h})$.\footnote{
    Expanding $\text{Var}(\bar{Y}_{j,h})$ via the law of total variance --- macro-level shocks are shared within a cell, while residuals average down as $\sigma_{res}^2/n_{j,h}$ --- gives $\text{Var}(\bar{Y}_{j,h}) = \sigma_{res}^2 \cdot \mathbb{E}[1/n_{j,h}] + \sigma_{macro}^2$.
    Substituting yields the full formula derived in \hyperref[app:bucket]{Appendix~\ref*{app:bucket}}: $\text{Var}(\hat{\tau}_{cell}) \approx \frac{4}{JH}\!\left[\sigma_{res}^2 \cdot \mathbb{E}[1/n_{j,h}] + \sigma_{macro}^2\right]$.
}
Crucially, however, the variance of a cell's outcome $\bar{Y}_{j,h}$ depends on its sample size. 
An unweighted estimator trades one penalty for another. On one hand, it shields the macro-level components from the $(1+cv^2)$ cluster imbalance penalty because each cell contributes equally to the estimate regardless of its sample size, preventing highly populated clusters from dominating the variance. 
On the other hand, it forces the estimator to place equal weight on highly volatile low-density cells, causing the residual variance component to inflate via $\mathbb{E}[1/n_{j,h}]$.

The theoretical choice between these estimators depends primarily on an asymptotic variance tradeoff: the $(1+cv^2)$ cluster imbalance penalty of the individual-level estimator versus Jensen's $\mathbb{E}[1/n_{j,h}]$ low-density penalty of the cell-level estimator. 
By equating their variances, we derive the exact analytical threshold ($cv^*$) at which the two estimators are equally efficient (\hyperref[app:bucket]{Appendix~\ref*{app:bucket}}):

\begin{equation}
cv^* = \sqrt{ \frac{S_{res}}{S_{macro}} \left( \mathbb{E}\left[\frac{1}{n_{j,h}}\right] - \frac{1}{\bar{n}} \right) }
\end{equation}

Consequently, severe cluster imbalance ($cv > cv^*$) asymptotically favors the cell-level estimator, whereas low cell density ($cv < cv^*$) strongly favors the individual-level estimator.

To quantify this tradeoff, recall our previous imbalanced marketplace scenario ($\bar{n} = 20$, $cv = 1.5$, $S_{res} = 80\%$, $S_{macro} = 20\%$). 
In such skewed distributions, Jensen's inequality is severe ($\mathbb{E}[1/n_{j,h}] \approx 0.16$). 
Evaluating the variance multipliers, the individual-level variance is proportional to $0.70$, whereas the cell-level variance is proportional to $0.33$. 
Here, the actual imbalance ($cv = 1.5$) far exceeds the theoretical threshold ($cv^* \approx 0.67$). 
Consequently, simply switching to a cell-level estimator cuts the total variance by more than half, effectively doubling statistical power without requiring additional data.

% \textcolor{blue}{
While the asymptotic variance tradeoff defining $cv^*$ is independent of the experiment's scale---as the number of clusters ($J$) and time periods ($H$) cancel out---statistical power is further governed by finite-sample inference. 
When an individual-level analysis size-weights imbalanced clusters, the variance becomes concentrated in a few large clusters, which reduces the effective degrees of freedom ($J_{\mathrm{eff}}$) for cluster-robust standard errors. 
In experiments with a small number of clusters ($J$), this reduction leads to wider critical $t$-values, effectively amplifying the cost of cluster imbalance. 
Conversely, cell-level analysis weights every cell equally, preventing any single cluster from dominating and thereby maintaining more stable degrees of freedom near $J-1$. 
Thus, finite-sample dynamics introduce an additional penalty on the individual-level estimator specifically in small, highly skewed experiments, reinforcing the asymptotic preference for cell-level aggregation in such regimes.
% } 
We detail these finite-sample considerations alongside our asymptotic derivations in \hyperref[app:bucket]{Appendix~\ref*{app:bucket}}.

\section{Conclusion}
We derived a closed-form asymptotic variance approximation for switchback experiments, eliminating the need for computationally expensive Monte Carlo simulations in power budgeting.
The formula reveals that cluster size imbalance ($cv$) imposes a permanent variance penalty on macro-level shocks.
Because this penalty does not shrink with data density ($\bar{n}$), practitioners cannot rely on large within-unit sample sizes to guarantee adequate statistical power.
To achieve reliable inference in imbalanced settings, experimenters must actively design around this constraint by increasing the number of independent randomizations ($J \times H$), balancing clusters, or employing variance reduction techniques.
Furthermore, our framework provides a rigorous heuristic for choosing between individual-level and cell-level estimators to preserve finite-sample power, and evaluates the limitations of advanced treatment assignment designs, proving mathematically that methods like stratification cannot eliminate the penalty on interaction variance. 
Finally, we demonstrate that variance reduction methods yield disproportionately greater efficiency gains when targeting spatial and spatiotemporal noise rather than residual noise.

\bibliography{references}

\newpage
\vspace{1cm}
\appendix
\noindent{\huge \textbf{Appendix}}

\section{Derivation of the Variance Formula}
\label{app:derivation}
We work with the model and estimator defined in Section~\ref{sec:model}.
The individual-level OLS point estimate $\hat{\tau}$ is mathematically equivalent to a ratio of the sums of outcomes over the sums of observations:
\begin{equation}
\hat{\tau}_{OLS} = \bar{Y}_{Treatment} - \bar{Y}_{Control} = \frac{\displaystyle\sum_{j,h} W_{j,h} \sum_{i=1}^{n_{j,h}} Y_{i,j,h}}{\displaystyle\sum_{j,h} W_{j,h}\, n_{j,h}}
- \frac{\displaystyle\sum_{j,h} (1 - W_{j,h}) \sum_{i=1}^{n_{j,h}} Y_{i,j,h}}{\displaystyle\sum_{j,h} (1 - W_{j,h})\, n_{j,h}}.
\end{equation}

By substituting the data generating process \eqref{eq:dgp} into the estimator, the grand mean $\mu$ cancels out exactly and the true treatment effect $\tau$ factors out.

\subsection{Linearizing via the Delta Method}
Both the numerators and denominators in $\hat{\tau}$ are random variables. 
We apply the Delta method (a first-order Taylor expansion around the expectations) to linearize the ratio. For a generic error component $S$ and denominator $N$, the approximation is $\frac{S}{N} \approx \frac{\mathbb{E}[S]}{\mathbb{E}[N]} + \frac{1}{\mathbb{E}[N]}(S - \mathbb{E}[S]) - \frac{\mathbb{E}[S]}{\mathbb{E}[N]^2}(N - \mathbb{E}[N])$.
Because the error components ($\alpha_j, \gamma_h, \delta_{j,h}, \epsilon_{i,j,h}$) have mean zero, the expected value of the numerator is $\mathbb{E}[S] = 0$. This simplifies the Taylor expansion to just $\frac{S}{\mathbb{E}[N]}$.

Given the 50/50 treatment probability, the expected total number of observations in either the treatment or control arm is $\mathbb{E}[N_1] = \mathbb{E}[N_0] = 0.5 J H \bar{n}$. Defining the treatment assignment indicator mapped to signs as $V_{j,h} = 2W_{j,h} - 1 \in \{-1, 1\}$, the linearized estimation error simplifies to:

\begin{equation}
\hat{\tau} - \tau \approx \frac{2}{J H \bar{n}} \sum_{j=1}^J \sum_{h=1}^H V_{j,h} \sum_{i=1}^{n_{j,h}} (\alpha_j + \gamma_h + \delta_{j,h} + \epsilon_{i,j,h})
\end{equation}

Because the treatment assignment $V_{j,h}$ is drawn independently across cells, this linearized error decomposes additively into four distinct components. 
Here, each $\Delta$ term represents the estimation error caused by that specific component---the difference between its weighted average in the treatment arm and its weighted average in the control arm:

\begin{equation}
\text{Var}(\hat{\tau}) \approx \text{Var}(\Delta_\alpha) + \text{Var}(\Delta_\gamma) + \text{Var}(\Delta_\delta) + \text{Var}(\Delta_\epsilon)
\end{equation}

We can now analyze the variance contribution of each error component individually.

\subsection{The Spatial Component ($\alpha_j$)}
The linearized error for the cluster component is:
\begin{equation}
\begin{aligned}
\Delta_\alpha &\approx \frac{2}{J H \bar{n}} \sum_{j=1}^J \sum_{h=1}^H V_{j,h} \sum_{i=1}^{n_{j,h}} \alpha_j \\
&= \frac{2}{J H \bar{n}} \sum_{j=1}^J \alpha_j \underbrace{\sum_{h=1}^H n_{j,h} V_{j,h}}_{U_j}
\end{aligned}
\end{equation}

where $U_j$ represents the net difference in sample size between treatment and control within cluster $j$. Because clusters are independent of each other, and $\alpha_j$ is independent of the assignments and cell sizes, the variance of their product is the product of their variances:

\begin{equation}
\text{Var}(\Delta_\alpha) = \left(\frac{2}{J H \bar{n}}\right)^2 J \sigma_{cl}^2 \text{Var}(U_j)
\end{equation}

To find $\text{Var}(U_j)$, we model the cell size $n_{j,h}$ as a random variable drawn from a distribution with a cluster-specific mean $m_j$. We use the Law of Total Variance, conditioning on this underlying true mean cell size for that cluster:

\begin{equation}
\text{Var}(U_j) = \mathbb{E}[\text{Var}(U_j \mid m_j)] + \text{Var}(\mathbb{E}[U_j \mid m_j])
\end{equation}

Because treatment $V_{j,h}$ is an independent coin flip with an expected value of 0, the expected imbalance in sample size between treatment and control given $m_j$ is exactly zero ($\mathbb{E}[U_j \mid m_j] = 0$). Since the expected value is a constant zero, its variance is zero, and the second term drops out entirely. 
For the first term, because cell sizes are conditionally independent across hours, the variance of the sum is the sum of the variances. 
The variance of $V_{j,h}$ is $1$, so $\text{Var}(n_{j,h} V_{j,h} \mid m_j) = \mathbb{E}[n_{j,h}^2 \mid m_j]$. 
Since $n_{j,h}$ follows a Poisson distribution with mean $m_j$, its second moment is $m_j + m_j^2$. 
Thus, $\text{Var}(U_j \mid m_j) = H (m_j + m_j^2)$. 
Note that this specific distributional assumption only dictates the vanishing $1/\bar{n}$ term in our final formula; the dominant $1+cv^2$ term arises strictly from the structural variance of the cluster means and is invariant to this choice.

Finally, we compute the expectation of this conditional variance over the clusters. 
The cluster mean sizes $m_j$ are drawn from an arbitrary distribution such that $\mathbb{E}[m_j] = \bar{n}$. 
Letting $cv$ denote the coefficient of variation of these mean cluster sizes, the variance is mathematically defined as $\text{Var}(m_j) = \bar{n}^2 cv^2$. 
Therefore, the expected second moment is $\mathbb{E}[m_j^2] = \bar{n}^2(1 + cv^2)$. 
Substituting this yields:

\begin{equation}
\text{Var}(U_j) = H \left( \bar{n} + \bar{n}^2(1 + cv^2) \right)
\end{equation}

Plugging this back into the cluster variance equation:

\begin{equation}
\text{Var}(\Delta_\alpha) \approx \frac{4 \sigma_{cl}^2}{J H} \left( \frac{1}{\bar{n}} + 1 + cv^2 \right)
\end{equation}

\subsection{The Temporal ($\gamma_h$) and Interaction ($\delta_{j,h}$) Components}
The temporal component follows an identical symmetric derivation. Grouping the imbalance by hour $h$ instead of cluster $j$, we obtain the exact same structure:
\begin{equation}
\text{Var}(\Delta_\gamma) \approx \frac{4 \sigma_{time}^2}{J H} \left( \frac{1}{\bar{n}} + 1 + cv^2 \right)
\end{equation}

The interaction component $\delta_{j,h}$ introduces potential temporal covariance because of underlying AR(1) dynamics. The variance of the sum over hours for a given cluster is:

\begin{equation}
\sum_{h=1}^H \text{Var}(\delta_{j,h} n_{j,h} V_{j,h}) + 2 \sum_{h < k} \text{Cov}(\delta_{j,h} n_{j,h} V_{j,h}, \, \delta_{j,k} n_{j,k} V_{j,k})
\end{equation}

Because randomization occurs independently at the cell level, $V_{j,h}$ and $V_{j,k}$ are independent mean-zero coin flips. Therefore, their cross-products have an expectation of zero, and the entire covariance term vanishes. This shows mathematically why high-frequency cell-level randomization completely neutralizes low-frequency AR(1) autocorrelation in the underlying data. The interaction variance thus simplifies to the same symmetrical form:

\begin{equation}
\text{Var}(\Delta_\delta) \approx \frac{4 \sigma_{int}^2}{J H} \left( \frac{1}{\bar{n}} + 1 + cv^2 \right)
\end{equation}

\subsection{The Residual Component ($\epsilon_{i,j,h}$) and Final Formula}
The residual errors are at the individual level and are not shared across a cluster or hour. They do not get multiplied by the cluster size terms that generate the $(1 + cv^2)$ penalty. The law of large numbers smooths them out cleanly:
\begin{equation}
\text{Var}(\Delta_\epsilon) \approx \frac{4 \sigma_{res}^2}{J H \bar{n}}
\end{equation}

By the Law of Total Variance, we sum the four independent components. Expressing each variance component as a share of the total variance $\sigma_{total}^2$ (e.g., $\sigma_{cl}^2 = S_{cl} \cdot \sigma_{total}^2$), we factor out the common terms to recover the closed-form approximation:
\begin{equation}
\text{Var}(\hat{\tau}) \approx \frac{4 \cdot \sigma_{total}^2}{J \cdot H} \left[ \frac{1}{\bar{n}} + S_{macro} \left( 1 + cv^2 \right) \right]
\end{equation}

\subsection{Decomposition into Naive A/B Variance and Switchback Penalty}
\label{app:decomp}
To explicitly connect this formula to standard experimental design, we can rewrite the expression to show how it relates to the variance of a naive, completely randomized A/B test. 

Starting from the expanded variance components, we have:
\begin{equation}
\text{Var}(\hat{\tau}) \approx \frac{4 \cdot \sigma_{res}^2}{J H \bar{n}} + \frac{4 \cdot \sigma_{macro}^2}{J H} \left(\frac{1}{\bar{n}} + 1 + cv^2\right)
\end{equation}

Distributing the macro term yields:
\begin{equation}
\text{Var}(\hat{\tau}) \approx \frac{4 \cdot \sigma_{res}^2}{J H \bar{n}} + \frac{4 \cdot \sigma_{macro}^2}{J H \bar{n}} + \frac{4 \cdot \sigma_{macro}^2}{J H} (1 + cv^2)
\end{equation}

Since the total number of observations in the experiment is $N = J \cdot H \cdot \bar{n}$, and the total variance is $\sigma_{total}^2 = \sigma_{res}^2 + \sigma_{macro}^2$, we can combine the first two terms:
\begin{equation}
\text{Var}(\hat{\tau}) \approx \frac{4 (\sigma_{res}^2 + \sigma_{macro}^2)}{N} + \frac{4 \cdot \sigma_{macro}^2}{J H} (1 + cv^2)
\end{equation}

This simplifies to:
\begin{equation}
\text{Var}(\hat{\tau}) \approx \underbrace{\frac{4 \cdot \sigma_{total}^2}{N}}_{\text{Naive A/B Variance}} + \underbrace{\frac{4 \cdot \sigma_{macro}^2}{J H} (1 + cv^2)}_{\text{Switchback Penalty}}
\end{equation}

This algebraic decomposition proves that the variance of a switchback experiment is equivalent to the variance of a standard A/B test plus a penalty term driven entirely by macro-level shocks and cluster size imbalance.

\section{Power and Minimum Detectable Effect}
\label{app:mde}

The variance approximation derived in \eqref{eq:variance} can be plugged directly into standard calculations for the minimum detectable effect (MDE) and sample-size budgeting.

\begin{equation}
\label{eq:mde}
\mathrm{MDE} = (z_{1-\alpha/2} + z_\beta) \times \sqrt{\text{Var}(\hat{\tau})}
\end{equation}

where $z$ denotes the standard normal quantiles for significance level $\alpha$ and power $1-\beta$.\footnote{
    The formulation in \eqref{eq:mde} assumes an asymptotic regime where the individual-level OLS estimator is normally distributed. 
    In practice, the specific standard error estimator used will dictate the appropriate test statistic. 
    In finite-sample regimes with a small number of clusters, practitioners typically pair cluster-robust variance estimators (e.g., CR2) with $t$-quantiles rather than standard normal $z$-quantiles.
    Crucially, because cluster sizes in marketplace settings are often highly skewed, the effective degrees of freedom for this $t$-distribution can be substantially smaller than $J-1$ \citep{carter2017}, necessitating data-driven degrees-of-freedom adjustments to maintain nominal coverage.
}

Substituting \eqref{eq:variance} into \eqref{eq:mde} yields the explicit formula for the minimum detectable effect:

\begin{equation}
\mathrm{MDE} \approx (z_{1-\alpha/2} + z_\beta) \sqrt{\frac{4 \sigma_{total}^2}{J H} \left[ \frac{S_{res}}{\bar{n}} + S_{macro} \left( \frac{1}{\bar{n}} + 1 + cv^2 \right) \right]}
\end{equation}

Alternatively, an experimenter can invert this relationship to calculate the required number of cell randomizations ($J \times H$) needed to achieve a desired power for a target effect size $\tau$:

\begin{equation}
J H \approx \frac{4 \sigma_{total}^2 (z_{1-\alpha/2} + z_\beta)^2}{\tau^2} \left[ \frac{S_{res}}{\bar{n}} + S_{macro} \left( \frac{1}{\bar{n}} + 1 + cv^2 \right) \right]
\end{equation}

\section{Finite Sample Limits: The Covariance Breakdown}
\label{app:boundary}
The first-order Delta method approximation systematically overpredicts the true variance of the estimator in highly skewed finite-sample regimes (such as when $cv \ge 2.0$ or $J \times H \le 10$). The intuition for this breakdown is revealed by extending the Taylor approximation to the second order.

\subsection{The 2nd-Order Taylor Expansion}
The OLS estimator's error term is structurally a fraction: $R = \frac{Y}{X}$, where $Y$ (the numerator) is the random sum of errors in the treated group, and $X$ (the denominator) is the random number of treated observations.
We expand $R$ around the expected values $\mu_Y = \mathbb{E}[Y]$ and $\mu_X = \mathbb{E}[X]$. Because the expected error of the experiment is zero, $\mu_Y = 0$. The second-order multivariate Taylor expansion is:

\begin{equation}
R \approx \frac{\mu_Y}{\mu_X} + \frac{1}{\mu_X}(Y - \mu_Y) - \frac{\mu_Y}{\mu_X^2}(X - \mu_X) + \frac{\mu_Y}{\mu_X^3}(X - \mu_X)^2 - \frac{1}{\mu_X^2}(Y - \mu_Y)(X - \mu_X)
\end{equation}

Because $\mu_Y = 0$, the first, third, and fourth terms vanish entirely, leaving:

\begin{equation}
R \approx \frac{Y}{\mu_X} - \frac{Y(X - \mu_X)}{\mu_X^2}
\end{equation}

\subsection{The Variance Correction}
To find the variance, we square this expansion (taking the expectation of $R$ as approximately 0):

\begin{equation}
R^2 \approx \left( \frac{Y}{\mu_X} \right)^2 - 2 \frac{Y^2(X - \mu_X)}{\mu_X^3} + \left( \frac{Y(X - \mu_X)}{\mu_X^2} \right)^2
\end{equation}

Taking the expectation yields the variance:

\begin{equation}
\text{Var}(R) \approx \frac{\mathbb{E}[Y^2]}{\mu_X^2} - 2 \frac{\mathbb{E}[Y^2(X - \mu_X)]}{\mu_X^3} + \frac{\mathbb{E}[Y^2(X - \mu_X)^2]}{\mu_X^4}
\end{equation}

The first term, $\frac{\mathbb{E}[Y^2]}{\mu_X^2}$, is precisely the first-order Delta method approximation that forms the basis of our analytical formula. 

The second term provides the crucial negative correction. The component $Y^2$ represents the squared error of the experiment, while $(X - \mu_X)$ represents the deviation of the treated sample size from its average.
In a highly skewed regime (e.g., $cv \ge 2.0$), if a disproportionately large cluster is assigned to treatment, the sample size is far above average ($X - \mu_X > 0$). Simultaneously, because this large cluster introduces substantial macro-level variance, the squared error $Y^2$ also increases substantially. Because $Y^2$ and $X$ increase together, they are strongly positively correlated, making the expectation $\mathbb{E}[Y^2(X - \mu_X)]$ a large positive value.

Because this positive expectation is multiplied by $-2 / \mu_X^3$, the second-order correction term is strictly negative. Therefore, the true variance of the estimator is naturally constrained, and the first-order approximation systematically serves as a conservative upper bound.

\section{Derivations for Advanced Assignment Designs}
\label{app:stratification}

A common heuristic to improve the efficiency of switchback experiments is to constrain the randomization using advanced operational designs. In this section, we formally derive the impact of within-cluster stratification (temporal blocking), pairing (cross-sectional matching), and mirroring (matched crossover) on the variance components.

Recall that in a completely unstratified (A/B) switchback, the individual-level estimator's variance is the sum of its independent components ($\Delta_\alpha, \Delta_\gamma, \Delta_\delta, \Delta_\epsilon$). The macro-level components each suffer from the $(1/\bar{n} + 1 + cv^2)$ penalty, while the residual component is scaled by $1/\bar{n}$.

\subsection{Variance under Within-Cluster Stratification}
Within-cluster stratification dictates that for every cluster $j$, treatment is balanced over time ($\sum_h V_{j,h} = 0$). 
This forces the expected difference in sample size between treatment and control for any cluster to zero. As a result, the variance contribution of the spatial component collapses from its amplified state down to its baseline:
\begin{equation}
\text{Var}_{strat}(\Delta_\alpha) \approx \frac{4 \sigma_{total}^2}{J H} \frac{S_{cl}}{\bar{n}}
\end{equation}

However, stratification does not balance assignments across clusters within a specific time period. The temporal ($\Delta_\gamma$) and interaction ($\Delta_\delta$) components remain fully penalized. The complete sum of variances under stratification is:
\begin{equation}
\text{Var}_{strat}(\hat{\tau}) \approx \frac{4 \sigma_{total}^2}{JH} \left[ \frac{S_{cl}}{\bar{n}} + S_{time}\left(\frac{1}{\bar{n}} + 1 + cv^2\right) + S_{int}\left(\frac{1}{\bar{n}} + 1 + cv^2\right) + \frac{S_{res}}{\bar{n}} \right]
\end{equation}

By expanding the terms inside the bracket and grouping those divided by $\bar{n}$, we obtain:
\begin{equation}
\text{Var}_{strat}(\hat{\tau}) \approx \frac{4 \sigma_{total}^2}{JH} \left[ \frac{1}{\bar{n}} (S_{cl} + S_{time} + S_{int} + S_{res}) + (S_{time} + S_{int})(1+cv^2) \right]
\end{equation}

Because the variance shares must sum to 1 by definition ($S_{cl} + S_{time} + S_{int} + S_{res} = 1$), the first term reduces to $\frac{1}{\bar{n}}$, yielding the final stratified variance:
\begin{equation}
\text{Var}_{strat}(\hat{\tau}) \approx \frac{4 \sigma_{total}^2}{J H} \left[ \frac{1}{\bar{n}} + (S_{time} + S_{int}) (1 + cv^2) \right]
\end{equation}

\subsection{Variance under Pairing and Mirroring}
To shield the temporal variance component ($S_{time}$) from the cluster imbalance penalty, experimental designers deploy \textit{pairing} (cross-sectional matching). By assigning opposite treatments to similarly sized clusters at the same time period, the expected sample sizes balance within every time period. This eliminates the quadratic dependency for the temporal component, collapsing it to its baseline:
\begin{equation}
\text{Var}_{paired}(\Delta_\gamma) \approx \frac{4 \sigma_{total}^2}{J H} \frac{S_{time}}{\bar{n}}
\end{equation}

By exact algebraic symmetry to stratification, grouping the $1/\bar{n}$ terms allows the variance shares to sum to 1, leaving the spatial and interaction components penalized:
\begin{equation}
\text{Var}_{paired}(\hat{\tau}) \approx \frac{4 \sigma_{total}^2}{JH} \left[ \frac{1}{\bar{n}} + (S_{cl} + S_{int})(1 + cv^2) \right]
\end{equation}

When pairing is combined with within-cluster stratification, the design is referred to as \textit{mirroring}. A mirrored design simultaneously balances expected treatment assignment across time within each cluster and across clusters within each time period, collapsing both $\text{Var}(\Delta_\alpha)$ and $\text{Var}(\Delta_\gamma)$ to their $1/\bar{n}$ baselines.

Summing the four components under a mirrored design gives:
\begin{equation}
\text{Var}_{mirror}(\hat{\tau}) \approx \frac{4 \sigma_{total}^2}{JH} \left[ \frac{S_{cl}}{\bar{n}} + \frac{S_{time}}{\bar{n}} + S_{int}\left(\frac{1}{\bar{n}} + 1 + cv^2\right) + \frac{S_{res}}{\bar{n}} \right]
\end{equation}

Grouping the $1/\bar{n}$ terms again yields the sum of all variance shares. The entire baseline variance fraction cleanly reduces to $\frac{1}{\bar{n}}$, leaving only the interaction penalty strictly isolated:
\begin{equation}
\text{Var}_{mirror}(\hat{\tau}) \approx \frac{4 \sigma_{total}^2}{J H} \left[ \frac{1}{\bar{n}} + S_{int}(1 + cv^2) \right]
\end{equation}

This reveals a fundamental limitation: because interaction shocks ($\delta_{j,h}$) are independent across clusters and time, no fixed assignment design can balance the interaction component in expectation. The $(1+cv^2)$ multiplier permanently attaches to the interaction variance share.

\section{Additional Simulation Results}
\label{app:sim_errors}
Table~\ref{tab:error_summary} summarizes the simulation validation results across all tested dimensions. The Delta method approximation performs exceptionally well across all standard settings (Safe zones) but transitions to a conservative upper bound in highly skewed or sparse-cluster Boundary regimes.

\begin{table}[H]
\centering
\caption{Mean and maximum approximation error across simulated regimes.}\label{tab:error_summary}
\renewcommand{\arraystretch}{1.30}
\resizebox{0.9\linewidth}{!}{%
\begin{tabular}{@{} l l c c c @{}}
    \toprule
    \textbf{Dimension} & \textbf{Zone} & \textbf{Regimes} & \textbf{Mean Error (\%)} & \textbf{Max Error (\%)} \\
    \midrule
    Number of clusters ($J$) & Interior & 5 & 6.57 & 14.28 \\
    Number of clusters ($J$) & Boundary & 1 & 18.29 & 18.29 \\
    Hours ($H$) & Interior & 5 & 3.92 & 9.51 \\
    Cluster size CV ($cv$) & Interior & 3 & 2.29 & 3.27 \\
    Cluster size CV ($cv$) & Boundary & 2 & 37.62 & 56.52 \\
    Mean cell density ($\bar{n}$) & Interior & 6 & 4.06 & 10.58 \\
    Residual variance share ($S_{res}$) & Interior & 4 & 2.47 & 4.19 \\
    Autocorrelation ($\rho$) & Interior & 4 & 2.35 & 4.13 \\
    Total variance scale & Interior & 3 & 1.78 & 2.19 \\
    \bottomrule
\end{tabular}}
\end{table}

\begin{figure}[htbp]
    \centering
    \includegraphics[width=0.9\textwidth]{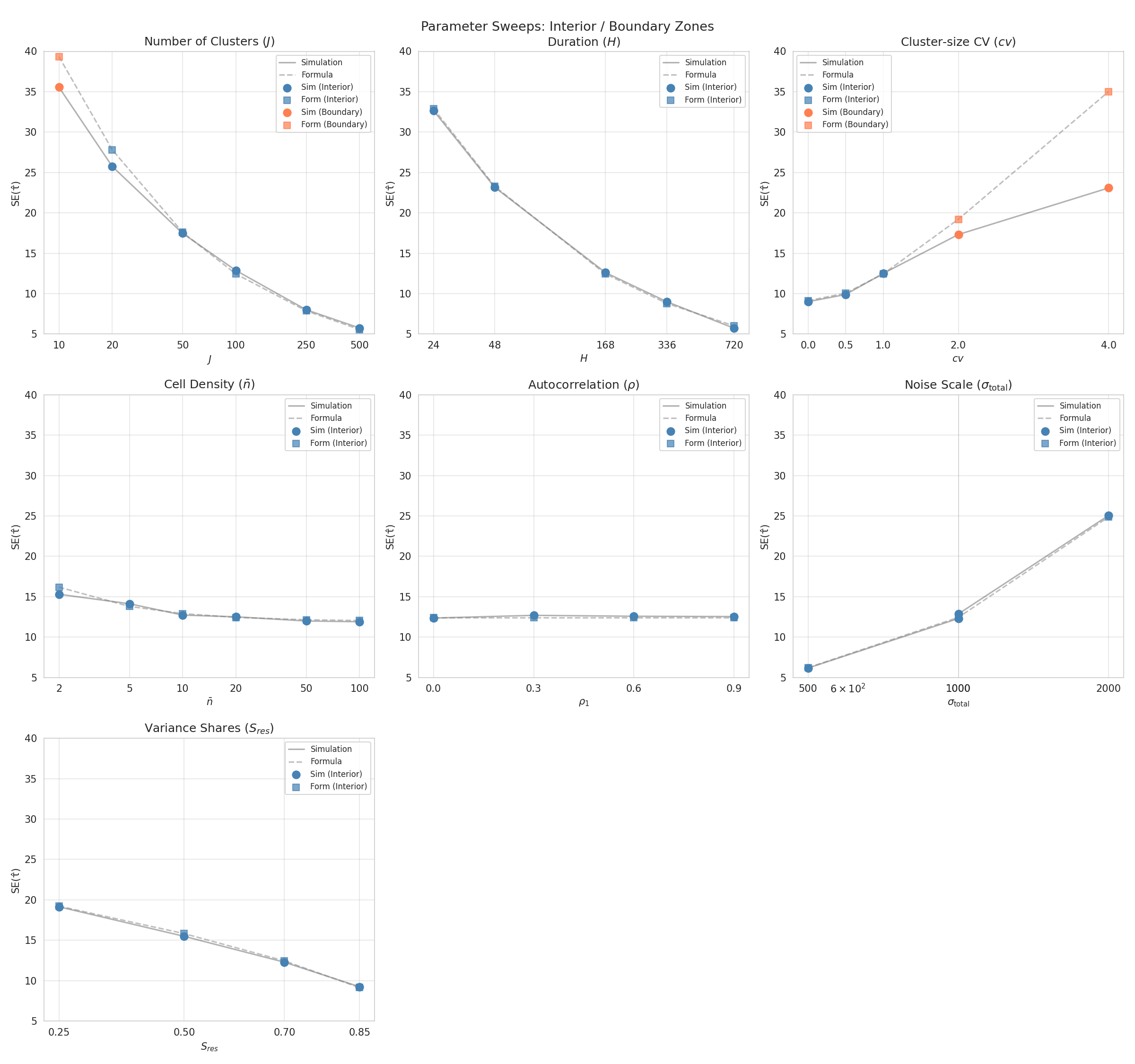}
    \caption{Theoretical versus Empirical Variance Across Simulation Parameter Regimes.}    \label{fig:ofat}
\end{figure}

Figure~\ref{fig:ofat} provides an overview of how the theoretical variance behaves compared to simulations along each independent dimension. 
Its first two panels demonstrate that the empirical variance decays exactly proportionally to $1/J$ and $1/H$, with the theoretical lines accurately bisecting the simulated points in the Interior regimes. 
Consistent with the theory, increasing cluster imbalance ($cv$) triggers an immediate convex surge in the variance. 

Further validating the structural mechanics, we find that the formula correctly captures the interplay of scaling laws and variance shares.
Scaling the overall baseline noise (Total Std) shifts the entire curve linearly, while varying the cell density ($\bar{n}$) confirms that the macro-variance floor limits efficiency gains.
Furthermore, shifting the variance budget from the residual component ($S_{res}$) to the macro-level components (holding the within-macro split proportional across $S_{cl}$, $S_{time}$, and $S_{int}$) dramatically inflates the total variance, confirming the multiplicative $(1+cv^2)$ penalty that applies uniformly to all macro-level shocks.
Finally, temporal autocorrelation drops out entirely when randomizing at the cell level, exactly as predicted. While temporal autocorrelation implies errors are highly correlated across time, treatment is assigned by independent coin flips at each cell. 
Thus, the correlated errors are randomly added and subtracted, and their covariance cancels out exactly in expectation. 
If treatment assignment were instead sticky or blocked over time, this autocorrelation would severely inflate the variance.

\section{Estimator Level Choice}
\label{app:bucket}
If an analyst aggregates data to the cell level, the outcome for cell $(j,h)$ is the simple average of its individual observations:

\begin{equation}
\bar{Y}_{j,h} = \mu + \tau W_{j,h} + \alpha_j + \gamma_h + \delta_{j,h} + \bar{\epsilon}_{j,h}
\end{equation}

where $\bar{\epsilon}_{j,h} = \frac{1}{n_{j,h}}\sum_i \epsilon_{i,j,h}$ is the average residual. Because the sum includes $n_{j,h}$ independent residuals, the variance of this average is $\sigma_{res}^2 / n_{j,h}$.
Taking the expectation over the cell sizes, the unconditional variance of a cell's outcome is:

\begin{equation}
\text{Var}(\bar{Y}_{j,h}) = \sigma_{res}^2 \cdot \mathbb{E}\left[\frac{1}{n_{j,h}}\right] + \sigma_{cl}^2 + \sigma_{time}^2 + \sigma_{int}^2
\end{equation}

An unweighted difference-in-means across these $J \times H$ cells is algebraically equivalent to a standard A/B test on $N = JH$ observations. Its variance is $\frac{4}{JH}\text{Var}(\bar{Y}_{j,h})$:

\begin{equation}
\text{Var}(\hat{\tau}_{cell}) \approx \frac{4}{J \cdot H} \left[ \sigma_{res}^2 \cdot \mathbb{E}\left[\frac{1}{n_{j,h}}\right] + \sigma_{macro}^2 \right]
\end{equation}

\subsection{Asymptotic Tradeoff and the Optimal Analysis Level}
Comparing this to our individual-level (size-weighted) formula reveals a strict structural tradeoff. The unweighted estimator protects macro-level shocks from the multiplicative $(1+cv^2)$ cluster imbalance penalty because it strips large clusters of their voting leverage. However, by assigning equal weight to low-density cells, it severely penalizes the idiosyncratic residual component. By Jensen's inequality, $\mathbb{E}[1/n_{j,h}] \gg 1/\bar{n}$, meaning volatile low-density cells will heavily inflate the total variance.

From a purely asymptotic perspective, we can analytically derive the exact parameter border that determines the optimal level of analysis. By equating the theoretical variance of the two estimators (omitting the common $\frac{4}{J\cdot H}$ constant), we find the threshold $cv^*$ at which the individual-level estimator and cell-level estimator are equally efficient:

\begin{equation}
\label{eq:boundary_raw}
\frac{S_{res}}{\bar{n}} + (1+cv^2)S_{macro} = S_{res} \mathbb{E}\left[\frac{1}{n_{j,h}}\right] + S_{macro}
\end{equation}

Solving for $cv$ and rewriting the terms using variance shares ($S_k = \sigma_k^2 / \sigma_{total}^2$) yields the theoretical boundary:

\begin{equation}
cv^* = \sqrt{ \frac{S_{res}}{S_{macro}} \left( \mathbb{E}\left[\frac{1}{n_{j,h}}\right] - \frac{1}{\bar{n}} \right) }
\end{equation}

This threshold formalizes the structural heuristics. 
Specifically, cell-level analysis ($cv > cv^*$) provides lower asymptotic variance in platforms with extreme cluster imbalance but uniformly large cells. 
Here, $\mathbb{E}[1/n_{j,h}] \approx 1/\bar{n}$, making the right-hand side of the threshold approach zero, meaning almost any meaningful cluster imbalance makes cell-level analysis preferable by shielding the estimator from the $(1+cv^2)$ penalty.
Conversely, individual-level analysis ($cv < cv^*$) is preferable if the marketplace has low cell density. 
In this regime, Jensen's inequality dictates that $\mathbb{E}[1/n_{j,h}] \gg 1/\bar{n}$. 
This substantial Jensen's gap highly inflates the threshold $cv^*$, making individual-level analysis vastly superior as the cell-level estimator's variance diverges.

\subsection{Finite-Sample Considerations}
The asymptotic logic above evaluates estimator efficiency under the assumption that critical values are drawn from a standard normal distribution. 
In practice, inference with cluster-robust standard errors relies on a $t$-distribution, where critical values depend on the effective degrees of freedom ($J_{\mathrm{eff}}$) \citep{imbens2016robust}.

When an individual-level analysis size-weights imbalanced clusters, the variance concentrates heavily in the largest clusters. 
As formalized by \citet{carter2017}, the effective degrees of freedom for a cluster-robust variance estimator (CRVE) degrade as the variance of the cluster sizes increases. 
Therefore, in highly skewed settings, the individual-level estimator exhibits reduced effective degrees of freedom ($J_{\mathrm{eff}} < J-1$).

In contrast, the cell-level estimator aggregates the data such that every cluster contributes exactly $H$ cells. 
Because the estimator is unweighted, large clusters contribute no more leverage to the macro-level variance than small clusters. 
While some heteroskedasticity remains due to varying cell sizes, the concentration of variance is substantially mitigated. 
This keeps the variance more evenly distributed across the $J$ clusters, protecting the estimator from degree of freedom degradation and maintaining $J_{\mathrm{eff}}$ closer to $J-1$. 
This approach parallels established cross-sectional econometrics practices, where aggregating to the group level provides a robust method for addressing small-sample clustering \citep{donald2007inference, cameron2015practitioner, abadie2017when}.

% \textcolor{blue}{
Consequently, finite-sample considerations introduce an amplifying penalty in small experiments that reinforces the asymptotic heuristics. 
Because the leading $\frac{4}{J \cdot H}$ term cancels out of the tradeoff boundary in \eqref{eq:boundary_raw}, the asymptotic threshold $cv^*$ depends only on the density and imbalance of the data, independent of the experiment's duration or the total number of clusters.
% } 
However, in finite samples, the total number of clusters ($J$) dictates the stability of the $t$-statistic. 
In large experiments (e.g., $J \gg 50$), $J_{\mathrm{eff}}$ remains sufficiently large that the $t$-statistic is stable, and the optimal estimator choice is determined almost entirely by the asymptotic variance threshold $cv^*$. 
However, in experiments with fewer clusters, the reduction in degrees of freedom associated with the individual-level estimator widens the critical $t$-values. 
This effectively increases the statistical penalty of cluster imbalance beyond the asymptotic $(1+cv^2)$ multiplier. 
As a result, in small, skewed experiments, cell-level analysis may yield greater statistical power even if the individual-level estimator offers a marginally tighter theoretical variance. 

% To assist practitioners in navigating this finite-sample tradeoff, we provide an open-source companion notebook alongside this paper. 
% The notebook acts as an interactive calculator that ingests pre-experimental sample data, computes exact finite-sample effective degrees of freedom for both estimators, applies the correct $t$-statistics, and outputs a concrete assessment of the most powerful analysis level.

\end{document}